\documentstyle[12pt]{article}
\def\s#1{{\small #1}}

\def\MC{Monte Carlo}

\def\EV{\s{EVENT}}
\def\bom#1{{\mbox{\boldmath $#1$}}}
\def\beq#1{\begin{equation}\label{#1}}
\def\beeq#1{\begin{eqnarray}\label{#1}}
\def\eeq{\end{equation}}
\def\eeeq{\end{eqnarray}}
\def\as{\alpha_S}
\def\asb{\bar\alpha_S}
\def\a0{\bar\alpha_0}

\def\ee{e^+e^-}
\def\Clim{\raisebox{-1ex}{\rlap{\tiny $\;\;C\to 0$}} 
\raisebox{0ex}
{$\;\;\;\,\sim\;\;\;\,$}}

\def\gtap{\raisebox{-.4ex}{\rlap{$\,\sim\,$}} 
\raisebox{.4ex}{$\,>\,$}}
\def\smax{{\mbox{\scriptsize max}}}
\def\pert{{\mbox{\scriptsize pert}}}

\def\Cpar{$C$-parameter}

\def\b0{\beta_0}

\def\frac#1#2{ {{#1} \over {#2} }}
\def\rat#1#2{\mbox{\small $\frac{#1}{#2}$}}
\def\half{\rat 1 2}
\def\thlf{\rat 3 2}
\def\thrq{\rat 3 4}

\def\naive{na\"{\i}ve}

\def\cav#1{Cambridge preprint Cavendish--HEP--#1}

\def\cpc#1#2#3{Computer Phys.\ Comm.\ #1 (19#3) #2}
\def\ib#1#2#3{ibid.\ #1 (19#3) #2}
\def\np#1#2#3{Nucl.\ Phys.\ B#1 (19#3) #2}
\def\pl#1#2#3{Phys.\ Lett.\ #1B (19#3) #2}
\def\pr#1#2#3{Phys.\ Rev.\ D #1 (19#3) #2}

\def\prl#1#2#3{Phys.\ Rev.\ Lett.\ #1 (19#3) #2}

\def\jhep#1#2#3{J.\ High Energy Phys.\ #1 (19#3) #2}

\def\zp#1#2#3{Z.\ Phys.\ C#1 (19#3) #2}

\begin{document}
\begin{titlepage}
\renewcommand{\thefootnote}{\fnsymbol{footnote}}
\begin{flushright}
     CERN-TH/98-14 \\
     Cavendish--HEP--97/16 \\
     hep-ph/9801350
\end{flushright}
\vspace*{\fill}
\begin{center}
{\Large \bf \boldmath Resummed $C$-Parameter Distribution\\[3mm]
in $\ee$ Annihilation\footnote{Research supported in part by
the U.K. Science and Engineering Research Council.}}
\end{center}
\par \vskip 2mm
\begin{center}
        {\bf S.\ Catani\footnote{On leave of absence from
        INFN, Sezione di Firenze, Italy.}} \\
        Theory Division, CERN, \\
        CH-1211 Geneva 23, Switzerland
        \par \noindent
        and
        \par \noindent
        {\bf B.R.\ Webber} \\
        Cavendish Laboratory, University of Cambridge,\\
        Madingley Road, Cambridge CB3 0HE, U.K.
\end{center}
\par \vskip 2mm
\begin{center} {\large \bf Abstract} \end{center}
\begin{quote}
We give perturbative predictions for the distribution of
the \Cpar\ event shape variable in $\ee$ annihilation,
including resummation of large logarithms in the
two-jet (small-$C$) region, matched to next-to-leading
order results.  We also estimate the leading non-perturbative
power correction and make a preliminary comparison with
experimental data.
\end{quote}
\vspace*{\fill}
\begin{flushleft}
     CERN-TH/98-14 \\
     Cavendish--HEP--97/16 \\
     January 1998
\end{flushleft} 
\end{titlepage}
\pagestyle{plain}
\renewcommand{\thefootnote}{\fnsymbol{footnote}}
\section{Introduction}\label{sec_intro}
The study of event shape distributions in $\ee$ annihilation
has proved to be a valuable testing-ground for QCD.
For an infrared-safe shape variable, i.e.\ one that is
insensitive to the splitting of a final-state momentum into 
collinear momenta and to the emission of soft momenta,
the distribution can be calculated order-by-order in
perturbation theory. However, one finds that higher-order
contributions are enhanced by large logarithmic factors
near the two-jet region, where most of the data reside.
Hence for a detailed quantitative understanding one should
resum as many of these terms as possible to all orders.
This has been achieved for a limited number of shape variables,
namely those shown to satisfy an {\em exponentiation} property,
to be specified in more detail later. These variables
include the thrust [\ref{CTTW}], heavy jet mass [\ref{hjm}],
jet broadening [\ref{broad}], energy-energy correlation [\ref{EEC}]
and differential two-jet rate [\ref{Rjets}].

In the present paper we show that the \Cpar\ [\ref{Cpar}]
shape variable also has the required property of exponentiation,
so that we are able to carry out the resummation of large
logarithms to the same accuracy as for those mentioned above.
The exponentiation of soft gluon contributions to QCD matrix elements
is a general phenomenon; the critical property for event shapes is
that the corresponding phase space should factorize with sufficient accuracy
to maintain exponentiation of at least the leading and next-to-leading
logarithms.  In the case of the \Cpar, we are able to show this by
exploiting a simple connection between the \Cpar\ and the thrust
in the dominant phase-space regions, namely those where $C$ is small.

In Sect.~\ref{sec_kin} we review the definition and kinematics of
the \Cpar, and the relationship between it and the thrust in the
small-$C$ region. Sect.~\ref{sec_FO} presents the fixed-order
predictions for the \Cpar\ distribution up to order $\as^2$,
together with the large logarithmic terms that appear at
small $C$. The resummation of these terms, to next-to-leading
logarithmic accuracy, is performed in Sect.~\ref{sec_resum}
using the connection with the thrust.

In order to describe the \Cpar\ distribution over the widest
possible range, one should match the resummed results to
those at fixed order outside the two-jet region. A suitable
matching procedure, chosen from those proposed in Ref.~[\ref{CTTW}],
is outlined in Sect.~\ref{sec_mat}.

A further improvement in the prediction can be made by
estimating non-perturbative effects, which have
been found to be substantial for many shape variables
at present energies. This is discussed in Sect.~\ref{sec_NP},
where we argue that the effects on the \Cpar\ distribution
can again be related to those for the thrust, via their
connection in the soft region.

Finally in Sect.~\ref{sec_data} we make a preliminary
comparison with experimental data and discuss the results.

\section{\boldmath Kinematics of the \Cpar}\label{sec_kin} 
The \Cpar\ was initially defined as [\ref{Cpar}]
\beq{Cdef}
C=3(\lambda_1\lambda_2+\lambda_2\lambda_3+\lambda_3\lambda_1)
\eeq
where $\lambda_\alpha$ $( 0 \leq \lambda_\alpha \leq 1 , \, \sum_\alpha
\lambda_\alpha = 1)$
are the eigenvalues of the linearized momentum tensor
\beq{ptens}
 \Theta^{\alpha \beta} = \frac{\sum_i {\bom p}^\alpha_i {\bom p}^\beta_i 
/|{\bom p}_i|}
{\sum_j |{\bom p}_j|}\; ,
\eeq
the sums being over all final-state particles. 
Neglecting particle
masses, we may express $C$ in terms of invariants as
\beq{Cinv}
C=3-\frac{3}{2}\sum_{i,j}\frac{(p_i\cdot 
p_j)^2}{(p_i\cdot Q)(p_j\cdot Q)}
\eeq
where $Q^{\mu}$ is the total four-momentum
\beq{Qdef}
Q^{\mu}=\sum_i p_i^{\mu}\; .
\eeq
Introducing the c.m.\ energy fractions $x_i=2(p_i\cdot 
Q)/Q^2$, we
can write Eq.~(\ref{Cinv}) as
\beq{Ca}
C = \rat 3 8 \sum_{i,j} 
 x_i x_j \sin^2\theta_{ij}\; .
\eeq

The kinematic range is $0 \leq C \leq 1$, with $C=0$ for a 
perfectly two-jet-like final state and 1 for an isotropic and
acoplanar distribution of final-state momenta.
In fact the maximal value $C=1$ can only be achieved 
when there are four or more final-state particles.
Planar events fill up the kinematic region $C \leq \thrq$.
For three particles, the maximum value 
$C=\thrq$ corresponds to the symmetric configuration
$x_1=x_2=x_3=\rat 2 3$.

\subsection{\boldmath \Cpar\ and thrust at small $C$}
Let us consider the definition (\ref{Cdef}) of the \Cpar\ and
assume the ordering of eigenvalues
$\lambda_1 \geq \lambda_2 \geq \lambda_3$.
It follows that 
in the small-$C$ limit the maximum eigenvalue approaches unity,
$\lambda_1 \to 1$, whilst $\lambda_2$ and $\lambda_3$ are parametrically
small, $\lambda_2 \sim \lambda_3 \sim {\cal O}(1-\lambda_1)$.
Using the normalization condition $\Sigma_\alpha \lambda_\alpha =1$, we
thus obtain
\beq{clambda}
C = 3 \{ (1-\lambda_1) - \half [(1-\lambda_1)^2 + \lambda_2^2
+ \lambda_3^2 ] \} \simeq 3 (1-\lambda_1) 
[1+ {\cal O}(1-\lambda_1) ] \;\;,
\eeq
or, equivalently,
\beq{clambda1}
C = 3 (1-\lambda_1) [1+ {\cal O}(C) ] \;, \;\;\;\;\; (C \to 0) \;.
\eeq

We can now relate the eigenvalue $\lambda_1$ to the thrust.
The definition of the thrust $T$ is [\ref{Farhi}]
\beq{tdef}
T = \mbox{Max} \frac{ \Sigma_i | {\bom p}_i \cdot 
{\bom n} |}{\Sigma_j | {\bom p}_j |} \;\;,
\eeq
where the maximum
is with respect to the unit 3-vector ${\bom n}$. At the maximum
${\bom n} = {\bom n}_T$, the thrust axis. Denoting by ${\bom n}_1$
the eigenvector of the linearized momentum tensor (\ref{ptens})
corresponding to the maximum eigenvalue $\lambda_1$,
by definition we have
\beq{lamax}
\lambda_1 = \frac{1}{\Sigma_j | {\bom p}_j |} 
\Sigma_i \frac{| {\bom p}_i \cdot 
{\bom n}_1 |^2}{|{\bom p}_i|} \leq
\frac{ \Sigma_i | {\bom p}_i \cdot 
{\bom n}_1 |}{\Sigma_j | {\bom p}_j |} \leq T\;\;,
\eeq
where the last inequality comes from Eq.~(\ref{tdef}).
From Eq.~(\ref{clambda1}) we thus obtain
\beq{cgeqt}
C \geq 3 (1-T) [1 + {\cal O}(C) ] \;, \;\;\;\;\; (C \to 0) \;.
\eeq

Equation (\ref{cgeqt}) implies that in the small-$C$ limit the
thrust approaches unity. The opposite is also true. In order
to show this, let us consider the separation of the final
state into two hemispheres $S_+$ and $S_-$ by the plane orthogonal
to the thrust axis ${\bom n}_T$. We can write Eq.~(\ref{Ca}) as
\beq{cpm}
C = \frac{3}{8} \left( \Sigma_{i,j}^{\prime} x_i x_j 
\sin^2 \theta_{ij} + \Sigma_{i,j}^{\prime \prime} x_i x_j 
\sin^2 \theta_{ij} \right) \;\;,
\eeq
where $\sum_{i,j}^{\prime}$ ($\sum_{i,j}^{\prime \prime}$)
denotes the sum over all final-state pairs belonging
to the same (opposite) hemisphere(s). Then, applying the
identity $\sin^2 \theta_{ij} = 2 (1 - \cos \theta_{ij}) -
(1 - \cos \theta_{ij})^2$ ($\sin^2 \theta_{ij} = 
2 (1 + \cos \theta_{ij}) - (1 + \cos \theta_{ij})^2$)
whenever the particles $i$ and $j$ belong to the same (opposite)
hemisphere(s) and using energy-momentum conservation,
it is straightforward to recast Eq.~(\ref{cpm}) in the form
\beq{cdt}
C = 3 \,[ \,2 (w_+ + w_-) - (w_+ - w_-)^2 ] - \Delta_T \;\;,
\eeq
where $w_+$ and $w_-$ are the rescaled invariant masses-squared
of the two hemispheres,
\beq{wpm}
w_{\pm} = \frac{1}{Q^2} 
\left( \sum_{i \in S_{\pm}} p_i \right)^2  \;\;,
\eeq
and $\Delta_T$ is defined by
\beq{deltat}
\Delta_T = \frac{3}{8} \left( \Sigma_{i,j}^{\prime} x_i x_j 
(1-\cos\theta_{ij})^2 + \Sigma_{i,j}^{\prime \prime} x_i x_j 
(1+\cos\theta_{ij})^2 \right) \;\;.
\eeq
Note that $\Delta_T$ is positive definite and that the expression in
the square bracket in Eq.~(\ref{cdt}) is exactly equal to $1-T^2$ 
[\ref{CTTW}]. Therefore from Eq.~(\ref{cdt}) we obtain
\beq{cleqt}
C = 3 (1-T^2) - \Delta_T \leq 3 (1-T^2) \leq 6(1-T)\;.
\eeq

Thus, combining the inequalities (\ref{cgeqt}) and (\ref{cleqt}),
we see that for sufficiently small $C$ (i.e., neglecting corrections
of order $C^2$) we have
\beq{Cbounds}
3(1-T) \leq C \leq 6(1-T)\;.
\eeq

\section{\boldmath The \Cpar\ in perturbation theory}\label{sec_FO}
\begin{figure}
  \centerline{
    \setlength{\unitlength}{1cm}
    \begin{picture}(0,7.5)
       \put(0,0){\includegraphics{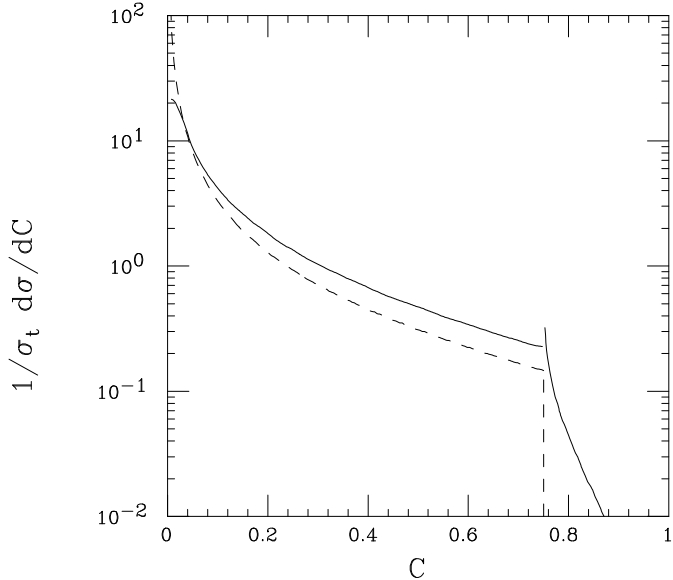}}
    \end{picture}}
  \caption[DATA]{
Fixed-order predictions of the \Cpar\ distribution for 
$\as = 0.12$. \\
Dashed: ${\cal O}(\as)$. Solid: ${\cal O}(\as^2)$.
 }\label{fig_Cfix}
\end{figure}

The \Cpar\ is manifestly insensitive to the splitting of a
final-state momentum into  collinear momenta ($\theta_{ij}=0$)
and to the emission of soft momenta ($x_i=0$). It follows that
the \Cpar\ distribution can be computed in QCD perturbation theory
as a power series expansion in the strong coupling $\as$.
For $C\neq 0$, the predicted distribution
has the general form
\beq{Cexp}
\frac{1}{\sigma_0} \frac{d\sigma}{dC}
 = \asb A(C) + \asb^2 B(C) + {\cal O}(\as^3)\; ,
\eeq
where $\asb=\as/2\pi$ and we normalize to the Born cross 
section $\sigma_0$, as was done in Refs.~[\ref{ERT},\ref{KN}].
The first-order distribution $A(C)$ 
is shown by the dashed curve in Fig.~\ref{fig_Cfix}.
Notice that it diverges as $C\to 0$; in fact
at small $C$ one finds
\beq{C3small}
A(C) = 4C_F \frac{1}{C}\left[\ln\left(\frac{6}{C}\right)
-\frac{3}{4}\right] + {\cal O}(\ln C) \;.
\eeq
At larger values of $C$, $A(C)$ smoothly approaches a finite
value at the three-parton upper limit $C=\thrq$:
\beq{C334}
A(\thrq) = \frac{256}{243}\pi\sqrt 3 C_F = 7.6433\;.
\eeq

It will be useful to define the fraction of events
with \Cpar\ values less than $C$:
\beq{Rdef}
R(C) = \int_0^C \frac{dC}{\sigma_t} \frac{d\sigma}{dC}
 = 1+ \asb R_1(C) + \asb^2 R_2(C) + {\cal O}(\as^3)
\eeq
where $\sigma_t$ is the total cross section.
Then since $R(C_\smax)=1$ and $\sigma_t/\sigma_0 =
1+\rat 3 2 C_F\asb +{\cal O}(\as^2)$, we have
\beeq{R12AB}
R_1(C) &=& -\int_C^{\thrq} A(C)dC \nonumber \\
R_2(C) &=& -\int_C^1 B(C)dC +\rat 3 2 C_F\int_C^{\thrq} 
A(C)dC\;.
\eeeq
Introducing
\beq{Ldef}
L \equiv \ln\left(\frac{6}{C}\right)
\eeq
we find as $C\to 0$
\beq{R1lim}
R_1(C) = -2C_F (L^2 - \rat 3 2 L -\rat 1 3 \pi^2 +\rat 5 
4) 
+ {\cal O}(C\ln C)\;.
\eeq
Notice that in first order we obtain up to two large logarithmic
factors $L$ at small $C$, corresponding to the emission of one
collinear and/or soft gluon. Note also that by rescaling $C$ by
a factor of 6 in the definition (\ref{Ldef})
we absorb all logarithms in Eq.~(\ref{R1lim}) into $L$.
In Sect.~\ref{sec_resum} we shall see that this is a consequence
of the relationship (\ref{cleqt}) between
the \Cpar\ and the thrust at small $C$.

Upon adding the second-order contribution $B(C)$, we 
obtain the solid curve in Fig.~\ref{fig_Cfix}, which exhibits
a number of new features [\ref{ERT}]: the behaviour at $C\to 0$
is modified, in fact becoming strongly negative divergent;
the distribution remains finite as $C\to \thrq^-$
but there is a new divergence as $C\to \thrq^+$; and
finally the distribution vanishes smoothly as $C\to 1$.

The new features at large $C$ arise from the abrupt change
in the ${\cal O}(\as)$ contribution $A(C)$ at $C=\thrq$,
see Eq.~(\ref{C334}). They have been discussed in
Ref.~[\ref{CW97}] and the resummation of large terms of the
type $\ln(C-\thrq)$ in this
region will be considered in a separate paper [\ref{CW98}].
In the present paper we concentrate on terms that diverge
as $C\to 0$, and the matching of the resummed and fixed-order
predictions in the region $0<C<\thrq$, where the latter remains
smooth.

In the region of small $C$,
the emission of two collinear and/or soft gluons can yield up to
four powers of $L=\ln (6/C)$. Thus, writing the 
logarithmic dependence explicitly, the second-order contribution
$R_2$ to the event fraction $R(C)$ defined in Eq.~(\ref{Rdef})
has the form
\beq{R2lim}
R_2(C) = \sum_{m=0}^{4} R_{2m} L^m + D_2(C)\;,
\eeq
where $D_2\to 0$ as $C\to 0$.
We have computed the coefficients $R_{24}, R_{23}, R_{22}$ analytically,
and find that their values, together with Eq.~(\ref{R1lim}), are consistent
with the following exponentiating form for $R(C)$:
\beq{Rexp}
R(C) = \left(1+\sum_{n=1}^\infty C_n \asb^n \right)
\exp\left(\sum_{n=1}^\infty \sum_{m=1}^{n+1} G_{nm} 
\asb^n L^m\right)
+\sum_{n=1}^\infty \asb^n D_n(C) \; .
\eeq
Here $C_n$ and $G_{nm}$ are $C$-independent coefficients and 
the remainder functions $D_n(C)$ 
vanish as $C\to 0$.  

In the next section we argue that the exponentiation formula
(\ref{Rexp}) is actually valid to all orders in perturbation theory.
Therefore it is convenient to express the coefficients in
Eq.~(\ref{R2lim}) in terms of those in Eq.~(\ref{Rexp}), as follows
\beeq{R2s}
R_{20}  & = & C_2 \;,\;\;\;\;
R_{21}  \> = \> G_{21} + C_1 G_{11} \;,\;\;\;\;
R_{22}  \> = \> G_{22} + \rat 1 2 G_{11}^2 + C_1 G_{12}\;,
\nonumber \\
R_{23}  & = & G_{23} + G_{12}G_{11} \;,\;\;\;\;
R_{24}  \> = \> \rat 1 2 G_{12}^2 \; .
\eeeq
The coefficients which determine
$R_{2m}$ for $m\geq 2$ are as given in Table~1.

\begin{table}
\begin{center}\begin{tabular}{|ccl|}\hline
 & &\\
$C_1$ &=& $+C_F(4\pi^2-15)/6$ \\
 & &\\
$G_{11}$ &=& $+3C_F$ \\
 & &\\
$G_{12}$ &=& $-2C_F$ \\
 & &\\
$G_{22}$ &=& $-C_F[48\pi^2C_F + (169-12\pi^2)C_A - 
22N_f]/36$ \\
 & &\\
$G_{23}$ &=& $-C_F(11C_A -2N_f)/3$ \\
 & &\\
\hline \end{tabular}\label{T1}
\caption{Coefficients $C_1$ and $G_{nm}$ in Eq.~(\ref{Rexp}).}
\vspace*{-0.25cm}
\end{center}
\end{table}

\begin{figure}
  \centerline{
    \setlength{\unitlength}{1cm}
    \begin{picture}(0,7.5)
       \put(0,0){\includegraphics{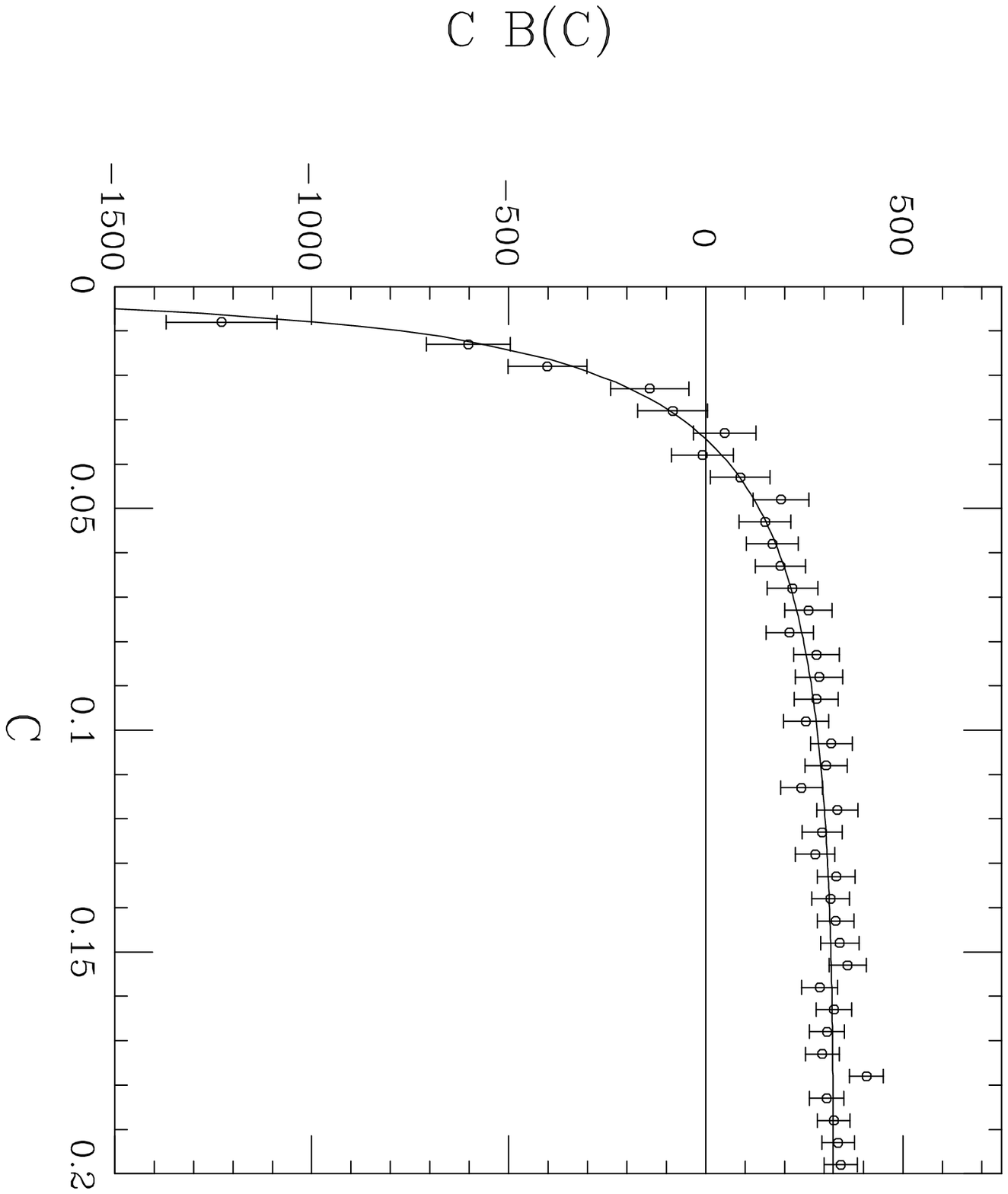}}
    \end{picture}}
  \caption[DATA]{
Second-order prediction of the \Cpar\ distribution for 
$C\to 0$. \\
Points: \EV\ \MC. Curve: Eqs.~(\ref{Bint},\ref{C2G21}).
 }\label{fig_Cpar_zero}
\end{figure}

The coefficients $C_2$ and $G_{21}$ which enter into
$R_{20}$ and $R_{21}$ are not known analytically but
can be fitted to numerical data from the \MC\ matrix
element evaluation program \EV\ [\ref{KN}]. From Eqs.~(\ref{R12AB})
and (\ref{Rexp}) we have, for $N_f=5$, 
\beeq{Bint}
-\int_C^1 dC B(C) &\!\!=\!\!& R_2(C) + \rat 3 2 C_F R_1(C) \\
&\!\!\Clim\!\!& 3.556 L^4 -20.889 L^3 -36.778 L^2 + 
(G_{21}+29.758)L + C_2+10.879\;.
\nonumber
\eeeq
By fitting we find
\beq{C2G21}
C_2 = 76.5\pm 2.9\;,\;\;\;\; G_{21} = 63.4\pm 6.0\;.
\eeq
Then Eq.~(\ref{Bint}) gives a good description of the 
matrix element, as illustrated in Fig.~\ref{fig_Cpar_zero}.
We checked that similar results are obtained with the
program \EV 2 [\ref{event2}].

\section{\boldmath Resummation for $C \to 0$}\label{sec_resum}
As explained in the Introduction, in order to carry out the
resummation of the $\log C$ contributions
for $C \to 0$, 
we exploit the connection between
the \Cpar\ and the thrust 
in the two-jet region, discussed in Sec.~\ref{sec_kin}.
However, to derive the relevant relation between $C$
and $1-T$, we cannot simply rely
on the kinematics but rather we have to consider QCD dynamics.

As discussed in detail in Ref.~[\ref{CTTW}], the structure of the
soft and collinear singularities in the multiparton QCD matrix elements
can be described in terms of a branching process. In the large-$T$ 
region and to next-to-leading logarithmic accuracy, this 
process takes place in an angular-ordered region around the
thrust axis. Because of this angular ordering (enforced by QCD
dynamics), we can introduce in Eq.~(\ref{cpm}) the approximation 
$\sin^2 \theta_{ij} \simeq 2 (1 - \cos \theta_{ij})$
($\sin^2 \theta_{ij} \simeq 2 (1 + \cos \theta_{ij})$)
in the terms appearing in $\sum_{i,j}^{\prime}$ 
($\sum_{i,j}^{\prime \prime}$). Equivalently, we can say
that when $T \to 1$ the phase-space region contributing
to $\Delta_T$ in Eq.~(\ref{cleqt}) is subdominant with respect
to that contributing to $3(1-T^2) \simeq 6(1-T)$. It follows that
to {\em next-to-leading logarithmic} accuracy at small $C$ we have
\beq{c1-t}
C \simeq 6 (1-T)\;.
\eeq 

In Ref.~[\ref{CTTW}] the perturbative expression for $R_T(\tau)$,
the fraction of events with thrust larger than $T=1-\tau$, was written
in the following form:
\beq{RTSig}
R_T(\tau=1-T) = K_T(\as)\Sigma_T(\tau,\as) + D_T(\tau,\as)\; .
\eeq
By definition the functions $K_T(\as), \,D_T(\tau,\as)$ and
$\Sigma_T(\tau,\as)$ are power series expansions in $\as$
whose coefficients are respectively constant in $\tau$,
vanishing for $\tau \to 0$ and polynomials in $L=\ln 1/\tau$.
The perturbative contributions to $R_T(\tau)$ which are 
logarithmically enhanced in the two-jet region $\tau =1-T \to 0$
are thus embedded in $\Sigma_T$. At small $\tau$ it becomes
important to resum the series of large logarithms in $\Sigma_T$.

By \naive\ counting of logarithms $R_T$, and hence $\Sigma_T$, contain
terms of the type $\as^n L^m$ with $m \leq 2n$. However, by explicitly
performing the resummation of the leading logarithms (those
with $n < m \leq 2n$), it was shown [\ref{CTTW}] that they {\em
exponentiate}. The word exponentiation means that the terms    
$\as^n L^m$ with $m > n+1$ are absent from $\ln R_T \,(\ln \Sigma_T)$,
although they do appear in $R_T$ itself. 

The result in Eq.~(\ref{c1-t}) allows us to obtain the leading and
next-to-leading $\log C$ contributions to the \Cpar\ distribution
simply by replacing $\ln(1/\tau)$ by $L=\ln (6/C)$ in the corresponding
formula for the thrust distribution. It follows that the function
$R(C)$ defined in Eq.~(\ref{Rdef}) has the same form as
Eq.~(\ref{RTSig}), that is,
\beq{RCSig}
R(C) = K(\as)\Sigma(C,\as) + D(C,\as)\; ,
\eeq
with
\beeq{lnsc}
\ln\Sigma (C,\as) &=& 
\sum_{n=1}^\infty\sum_{m=1}^{n+1} G_{nm} \asb^n L^m \\
& = & L \,g_1(\as L) + g_2(\as L) + \as \,g_3(\as L) + 
\cdots \nonumber \;.
\eeeq
The function $g_1$
resums all the leading logarithmic contributions $\as^n L^{n+1}$,
while $g_2$ contains the next-to-leading logarithmic terms
$\as^n L^n$,  and $g_3$ etc.\ represent the remaining subdominant
corrections $\as^n L^m$ with $0<m<n$. All the functions $g_i$ vanish at
$L =0$ since they resum terms with $m>0$.

Because of the result in Eq.~(\ref{c1-t}), the 
functions $g_1$ and $g_2$ in Eq.~(\ref{lnsc}) are the same as those
for the thrust distribution, given in Ref.~[\ref{CTTW}],
provided we identify $L$ with $\ln (6/C)$. 
This relation between $\ln \Sigma(C)$ and 
$\ln \Sigma_T(\tau)$ breaks down at subdominant orders, i.e.\ the 
functions $g_i$ for $i>2$ are different in the two cases.
Similarly the non-logarithmic coefficient functions $K$ and $K_T$
and the remainder functions $D$ and $D_T$ are different,
as may be seen by comparing the fixed-order results in
Sec.~\ref{sec_FO} with the corresponding ones in Ref.~[\ref{CTTW}].

\section{Matching with fixed order}\label{sec_mat}
To combine the fixed-order and resummed predictions of the previous
two sections without any double counting, we may adopt any of the
matching procedures proposed in Ref.~[\ref{CTTW}]. In the so-called
log-$R$ matching scheme, for example, one writes the next-to-leading
order expression (\ref{Rdef}) in the equivalent form
\beq{Rfexp}
\ln R(C) = \asb R_1(C) + \asb^2\{R_2(C)-\half
[R_1(C)]^2\} + {\cal O}(\as^3)
\eeq
and then replaces the computed logarithmic terms by the
corresponding resummed expressions, to obtain
\beeq{Rlmat}
\ln R(C) & = & L \,g_1(\as L) + g_2(\as L)
+\asb\{R_1(C)-G_{11}L -G_{12}L^2\} \nonumber \\
&& + \asb^2\{R_2(C)-\half [R_1(C)]^2-G_{22}L^2 -G_{23}L^3\}\;.
\eeeq 
An advantage of this scheme is that, since all terms
are exponentiated, it is not necessary to separate out the
coefficient and remainder functions $K$ and $D$ in
Eq.~(\ref{RCSig}) explicitly.
A potential disadvantage is that the resummed terms may
dominate over the fixed-order ones far from the two-jet
region, if the latter become very small there.  This
should not be a severe problem for the \Cpar\ distribution,
since the first-order prediction does not vanish at the
three-parton boundary, Eq.~(\ref{C334}).
Thus we expect the log-$R$ prescription (\ref{Rlmat})
to be satisfactory at high energies, as long as matching
is limited to the region $C<\thrq$. Note, however, that
the sensitivity to the matching procedure can increase at
low energies, and therefore other schemes should also be
tried.

\section{Non-perturbative power corrections}\label{sec_NP}
Even at high energies ($Q\gtap M_Z$), it has been found that
substantial non-perturbative `hadronization corrections' need
to be applied to perturbative predictions concerning event
shape variables. Traditionally these have been estimated
[\ref{resexp},\ref{SLD}]
using \MC\ hadronization models [\ref{JS},\ref{HW}].
Recently, an alternative `renormalon' or `dispersive'
method for estimating them has been developed
[\ref{Web94}-\ref{DasWebshap}].  According
to this approach, the dominant non-perturbative
correction to the thrust distribution, at thrust
values that are not too close to the kinematic
boundaries, should correspond to a simple
shift in the resummed perturbative prediction,
of the form [\ref{DokWeb97}]
\beq{RTpow}
R_T(\tau) = R_T^\pert (\tau-4 A_1/Q)\;,
\eeq
where $R_T^\pert$ is the function called $R_T$ in Eq.~(\ref{RTSig})
and $A_1$ is a non-perturbative parameter
to be determined experimentally.

Because of the close connection (\ref{Cbounds})
between the \Cpar\ and thrust, the same type of
result holds for the \Cpar\ distribution.
Furthermore, the assumption that the leading
non-perturbative effect is associated with a
universal effective strong coupling at low
scales [\ref{DokMarWeb}] leads to a relationship
between the shifts in the two distributions.
Consider the emission of a single soft
gluon at angle $\theta$ with energy fraction $x$.
We have from Eq.~(\ref{Ca}) a contribution to $C$
of $\thlf x \sin^2\theta$,
whereas the contribution to $\tau$ is
$\half x \min\{1-\cos\theta,1+\cos\theta\}$.
As expected, the ratio is between 3
(at $\theta=\pi/2$) and 6 (in the collinear
regions, $\theta=0,\pi$). Integrating at a fixed small value
of the gluon transverse momentum $k_t$, we obtain 
\beq{delCTeqn}
\left\{\begin{array}{c}\delta C\\ \delta\tau\end{array}\right\}
= 2C_F\frac{\as(k_t)}{\pi}\int_0^1\frac{dx}{x}\int_0^\pi
\frac{d\theta}{\sin\theta}\delta(k_t-\half x Q\sin\theta)
\left\{\begin{array}{c}\thlf x \sin^2\theta\\
\half x \min\{1-\cos\theta,1+\cos\theta\}\end{array}\right\}\;,
\eeq
giving [\ref{Web94}]
\beq{delCTrat}
\frac{\delta C}{\delta\tau} = \frac{3\pi}{2}\;.
\eeq
This is smaller than the factor of 6 in Eq.~(\ref{c1-t}),
because the non-perturbative effect comes from the
whole soft region of gluon emission, whereas the
perturbative logarithmic enhancement comes from
the collinear region. As in the case of the thrust
distribution, the soft-gluon contribution exponentiates
and, taking into account Eq.~(\ref{delCTrat}), we have
\beq{Rpow}
R(C) = R^\pert (C-6\pi A_1/Q)\;.
\eeq

In Ref.~[\ref{NaSey}] it was pointed out that higher-loop effects
can alter the coefficients of the power corrections to event
shapes predicted by the dispersive approach. However,
recent studies of two-loop contributions [\ref{Milan}]
have shown that Eqs.~(\ref{RTpow}) and (\ref{Rpow}) remain
valid, provided that the relationship between
the non-perturbative constant $A_1$ and the effective
coupling at low scales is renormalized
by a common overall `Milan factor' ${\cal M}\simeq 1.8$.

\section{Comparison with data}\label{sec_data}
\begin{figure}
  \centerline{
    \setlength{\unitlength}{1cm}
    \begin{picture}(0,7.5)
       \put(0,0){\includegraphics{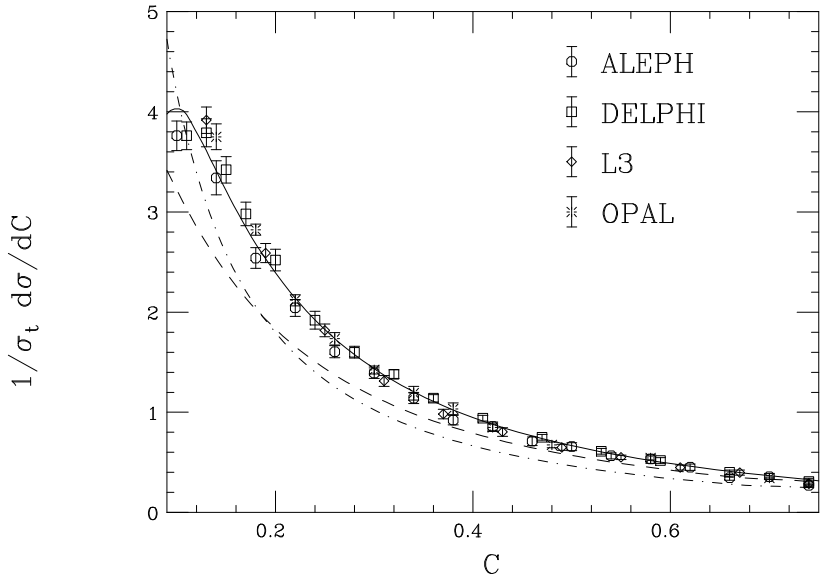}}
    \end{picture}}
  \caption[DATA]{
The \Cpar\ distribution at $Q=M_Z$. Solid curve: Eq.~(\ref{Rpow})
with $A_1=0.24$ GeV. Dashed: $A_1=0$. Dot-dashed: ${\cal O}(\as^2)$.
 }\label{fig_cshape}
\end{figure}

At present the experimental data on the \Cpar\ distribution
are limited to high energies, $Q=M_Z$ [\ref{SLD},\ref{expL1}]
and above [\ref{expL2}]. Figure \ref{fig_cshape} shows the
LEP1 data [\ref{expL1}] together with various theoretical
calculations, all assuming $\as(M_Z)=0.118$. The solid curve
represents Eq.~(\ref{Rpow}) with the best-fit value of $A_1=0.24$ GeV,
which gives very good agreement throughout the region $0.09<C<0.75$.
For the function $R^\pert$ the resummed expression with log-$R$
matching, Eq.~(\ref{Rlmat}), was used. The dashed curve shows the
effect of neglecting the non-perturbative shift $A_1$, while the
dot-dashed curve shows the fixed-order prediction (\ref{Cexp}),
with neither resummation nor non-perturbative correction.
 
The corresponding value of $A_1$ obtained from fitting the
thrust distribution [\ref{DokWeb97}] is $A_1=0.22$ GeV.
Thus the results obtained from the \Cpar\ and thrust
data are consistent within 10\%.
Taking into account the Milan enhancement factor ${\cal M}$
in the coefficient of the power correction
[\ref{Milan}], these results correspond to a
value $\a0\simeq 0.35$ for the non-perturbative parameter
$\a0$ (the mean value of the effective strong coupling
at scales below 2 GeV) introduced in
Refs.~[\ref{DokWeb95},\ref{DokWeb97}].

The fitted values of $\as$ and $A_1$ are not well constrained separately
by the existing \Cpar\ data, owing to the lack of measurements of
this quantity at lower energies. A comprehensive analysis over a wide
range of energies, similar to those presented in
Refs.~[\ref{DELPHI}-\ref{ALEPH}] for other shape variables,
would therefore be very useful.

\section*{Acknowledgements}
We are most grateful for comments and encouragement from
S.\ Kluth. BRW thanks CERN Theory Division for hospitality while
part of this work was carried out.

\newpage
\section*{References}
\begin{enumerate}
\item  \label{CTTW}
       S.\ Catani, L.\ Trentadue, G.\ Turnock and B.R.\ Webber,
       \pl{263}{491}{91}; \np{407}{3}{93}.
\item  \label{hjm}
       S.\ Catani, G.\ Turnock and B.R.\ Webber, 
       \pl{272}{368}{91}.
\item  \label{broad}
       S.\ Catani, G.\ Turnock and B.R.\ Webber, 
       \pl{295}{269}{92}.
\item  \label{EEC}
       J.\ Kodaira and L.\ Trentadue, \pl{112}{66}{82},
       \pl{123}{335}{83};\\ 
       J.C.\ Collins and D.E.\ Soper, \np{197}{446}{82};\\
       G.\ Turnock, \cav{92/3},
       Ph.D.\ Thesis, University of Cambridge, 1992.
\item  \label{Rjets}
       S.\ Catani, Yu.L.\ Dokshitzer, M.\ Olsson, G.\ Turnock
       and B.R.\ Webber, \pl{269}{432}{91}; \\
       G.\ Dissertori and M.\ Schmelling, \pl{361}{167}{95}.
\item  \label{Cpar}
       G.\ Parisi, \pl{74}{65}{78};\\
       J.F.\ Donoghue, F.E.\ Low and S.Y.\ Pi, \pr{20}{2759}{79}.
\item  \label{Farhi}
       E.\ Farhi, \prl{39}{1587}{77}.
\item  \label{ERT}
       R.K.\ Ellis, D.A.\ Ross and A.E.\ Terrano, \np{178}{421}{81}. 
\item  \label{KN}
       Z.\ Kunszt, P.\ Nason, G.\ Marchesini and B.R.\ Webber,
       in `Z Physics at LEP 1', CERN 89-08, vol.~1, p.~373.
\item  \label{CW97}
       S.\ Catani and B.R.\ Webber, \jhep{10}{5}{97} (hep-ph/9710333). 
\item  \label{CW98}
       S.\ Catani and B.R.\ Webber, in preparation.
\item  \label{event2}
       S.\ Catani and M.H.\ Seymour, \pl{378}{287}{96} (hep-ph/9602277),
       \np{485}{291}{97} (hep-ph/9605323).
\item  \label{resexp}
       ALEPH Collaboration, D.\ Decamp et al., \pl{284}{163}{92};\\
       L3 Collaboration, O.\ Adriani et al., \pl{284}{471}{92};\\
       OPAL Collaboration, P.D.\ Acton et al., \zp{59}{1}{93},
       R.\ Akers et al., \zp{68}{519}{95};\\
       DELPHI Collaboration, P.\ Abreu et al., \zp{59}{21}{93};\\
       TOPAZ Collaboration, Y.\ Ohnishi et al., \pl{313}{475}{93}.
\item  \label{SLD}
       SLD Collaboration, K.\ Abe et al., \pr{51}{962}{95} (hep-ex/9501003).
\item  \label{JS}
       T.\ Sj\"ostrand, \cpc{82}{74}{94} (hep-ph/9508391). 
\item  \label{HW}
       G.\ Marchesini, B.R.\ Webber, G.\ Abbiendi,
       I.G.\ Knowles, M.H.\ Seymour, L.\ Stanco,
       \cpc{67}{465}{92} (hep-ph/9607393). 
\item  \label{Web94}
       B.R.\ Webber, \pl{339}{148}{94} (hep-ph/9408222);
       see also {\em Proc.\ Summer School on Hadronic Aspects of
       Collider Physics, Zuoz, Switzerland, 1994} (hep-ph/9411384).
\item  \label{KorSte}
G.P.\ Korchemsky and G.\ Sterman, \np{437}{415}{95} (hep-ph/9411211),
in Proc.\ 30th Rencontres de Moriond,
Meribel les Allues, France, March 1995 (hep-ph/9505391); 
G.P.\ Korchemsky, G.\ Oderda and G.\ Sterman, Stony Brook preprint
ITP-SB-97-41, to appear in Proc.\ DIS97 (hep-ph/9708346).
\item  \label{BenBra} 
        M.\ Beneke and V.M.\ Braun, \np{454}{253}{95} (hep-ph/9506452). 
\item  \label{DokWeb95}
       Yu.L.\ Dokshitzer and B.R.\ Webber, \pl{352}{451}{95}
       (hep-ph/9504219).
\item  \label{AkZak}
R.\ Akhoury and V.I.\ Zakharov, \pl{357}{646}{95} (hep-ph/9504248), 
\np{465}{295}{96} (hep-ph/9507253).
\item  \label{NaSey}
       P.\ Nason and M.H.\ Seymour, \np{454}{291}{95} (hep-ph/9506317).
\item  \label{DokMarWeb}
       Yu.L.\ Dokshitzer, G.\ Marchesini and B.R.\ Webber,
       \np{469}{93}{96} (hep-ph/9512336).
\item  \label{DokWeb97}
       Yu.L.\ Dokshitzer and B.R.\ Webber, \pl{404}{321}{97}
       (hep-ph/9704298).
\item  \label{DasWebshap}
       M. Dasgupta and B.R. Webber, Cav\-end\-ish-HEP-96/5,
       to appear in Z.\ Phys.\ C (hep-ph/9704297).
\item  \label{Milan}
       Yu.L.\ Dokshitzer, A.\ Lucenti, G.\ Marchesini and G.P.\ Salam,
       Milan preprints IFUM-573-FT (hep-ph/9707532), IFUM-601-FT.
\item  \label{expL1}
       DELPHI Collaboration, P.\ Abreu et al., \zp{54}{55}{92};\\
       OPAL Collaboration, P.D.\ Acton et al., \ib{C55}{1}{92};\\
       L3 Collaboration, B.\ Adeva et al., \ib{C55}{39}{92};\\
       ALEPH Collaboration, D.\ Buskulic et al., \ib{C55}{209}{92}.
\item  \label{expL2}
       OPAL Collaboration, G.\ Alexander et al., \zp{72}{191}{96};
       K. Ackerstaff et al., \ib{C75}{193}{97}.
\item  \label{DELPHI}
       DELPHI Collaboration, P.\ Abreu et al., \zp{73}{229}{97};
       D.\ Wicke, in Proc.\ QCD97, Montpellier, July 1997
       (hep-ph/9708467).
\item  \label{JADEOPAL}
       JADE Collaboration, P.A.\ Movilla Fernandez et al.,
       Aachen preprint PITHA-97-27 (hep-ex/9708034); 
       O.\ Biebel, Aachen preprint PITHA-97-32 (hep-ex/9708036). 
\item  \label{ALEPH}
       ALEPH Collaboration, contributed paper LP-258 at Int.\ Symposium
on Lepton Photon Interactions, Hamburg, July 1997.
\end{enumerate}
\end{document}